\documentclass[runningheads]{llncs}
\usepackage[T1]{fontenc}
\usepackage{amsmath}
\usepackage{graphicx}
\usepackage{xcolor}
\usepackage{caption}
\usepackage{subcaption}
\usepackage{multirow}
\usepackage{cite}
\begin{document}
\title{Date-Driven Approach for Identifying State of Hemodialysis Fistulas: Entropy-Complexity and Formal Concept Analysis}

% \thanks{Supported by organization x.}

\titlerunning{Date-Driven Approach for Identifying State of Hemodialysis Fistulas}

\author{Vasilii A. Gromov\inst{1}\orcidID{0000-0001-5891-6597} \and E.I. Zvorykina \inst{1}\orcidID{0000-0001-6085-2051} \and Yurii N. Beschastnov \inst{1}\orcidID{0000-0001-6511-5894} \and Majid Sohrabi\inst{1,2}\orcidID{0000-0003-3695-604X}}
\authorrunning{Gromov et al.}
\institute{School of Data Analysis and Artificial Intelligence, Faculty of Computer Science, HSE University, 109028, Moscow, Russian Federation
\and
Laboratory for Models and Methods of Computational Pragmatics, HSE University, Moscow, Russian Federation
\\ \email{(stroller@rambler.ru)}\\
\email{(y.zvorykina@gmail.com)}\\
\email{(y.beschastnov@mail.ru)}\\
\email{(msohrabi@hse.ru)}\\}
\maketitle              % typeset the header of the contribution
\begin{abstract}
The paper explores mathematical methods that differentiate regular and chaotic time series, specifically for identifying pathological fistulas. It proposes a noise-resistant method for classifying responding rows of normally and pathologically functioning fistulas. This approach is grounded in the hypothesis that laminar blood flow signifies normal function, while turbulent flow indicates pathology. The study explores two distinct methods for distinguishing chaotic from regular time series. The first method involves mapping the time series onto the entropy-complexity plane and subsequently comparing it to established clusters. The second method, introduced by the authors, constructs a concepts-objects graph using formal concept analysis. Both of these methods exhibit high efficiency in determining the state of the fistula.

\end{abstract}
\section{Introduction}
The escalating prevalence of chronic kidney disease (CKD) continues to grow annually and has already reached levels comparable to other socially significant conditions, such as hypertension, diabetes, obesity, and metabolic syndrome ~\cite{ref1,ref2}. The conventional approach for patients requiring hemodialysis involves the establishment of permanent vascular access in the form of an arteriovenous fistula. However, this method is plagued by frequent occurrences of thrombosis, thereby jeopardizing patient safety. Currently, the global incidence of end-stage chronic kidney failure has risen to approximately $0.1\%$ \cite{ref2,ref3}. The advent of the SARS-CoV-2 pandemic has significantly impacted these statistics, as acute kidney injury has emerged as one of the most common extrapulmonary manifestations of COVID-19, often associated with a considerable decline in patient prognosis and outcomes \cite{ref1}.  In a majority of cases~\cite{ref4}, the sole recourse for preserving a patient's life lies in the diligent management of an arteriovenous fistula through hemodialysis procedures. Numerous studies underscore the advantages of native arteriovenous fistulas (AVF) over alternative approaches~\cite{ref5,ref6}. The timely detection of improper fistula function becomes challenging when reliant on patient self-monitoring, which has fueled an intensified interest in automated methods for identifying malfunctioning fistulas. Within this problem domain, the bruit series of the fistula~\cite{ref7} stands out as the most effective source of information. From a mathematical standpoint, this problem represents the resolution of an inverse problem employing some form of intelligent system~\cite{ref8}-\cite{ref10}. In this context, the direct problem involves a complex system characterizing the dynamics of blood flow within the human body.
This paper presents a methodology for automatically distinguishing between series corresponding to normally functioning fistulas and those exhibiting pathological behavior. The underlying hypothesis postulates laminar blood flow in normally functioning fistulas and turbulent flow in pathologically functioning ones~\cite{ref11}-\cite{ref13}, which can be attributed to stenosis and thrombosis of the blood vessels~\cite{ref14}.

The rest of the paper is organized as follows. In the second section, we discuss related works. In the third section, we introduce the proposed algorithm in detail. In the fourth section, we analyze the results of algorithm evaluation on data. Finally, in the Conclusion section, we provide the conclusion and potential applications of the proposed methodology.

\section{Related works}
The first documented instances of numerical analysis of AVF bruit date back to the mid-1980s and highlighted the correlation between sound amplitude variations and AVF functioning states. Subsequently, numerical analysis of phono-angiography signals has been primarily based on a two-class classification scheme, distinguishing AVF into "good" and "bad" categories, with only a few authors exploring a multi-class assessment approach, such as categorization into "good," "medium," and "bad" \cite{ref15}. Recent strides have been made towards developing technical solutions for continuous AVF bruit monitoring and diagnostics \cite{ref16}-\cite{ref17}. Nevertheless, the demand for a cost-effective, automated diagnostic device for assessing AVF status remains unmet.

From a mathematical perspective, this task revolves around differentiating between regular and chaotic time series. Technically, implementing the methodology is straightforward, necessitating only the recording of fistula sounds with a mobile phone and their subsequent automatic processing using a dedicated application. Two methods for distinguishing chaotic series from regular ones have been considered. The first involves determining the series' position on the "entropy-complexity" plane and comparing it with identified clusters of values derived from the set of time series. The method proposed by the authors entails constructing an objects-concepts graph \cite{ref18} within the framework of formal concept analysis \cite{ref19} -\cite{ref21} based on the examined time series and comparing its characteristics with those of graphs definitively associated with specified types of time series.

The methodology we propose hinges on distinguishing between chaotic and regular time series. The problem at hand—differentiating between chaotic and regular time series of blood flow in an arteriovenous fistula—possesses two features, one simplifying the task and the other complicating it. Firstly, during the observation periods, the series are generated by the same dynamical system with consistent parameter values \cite{ref22} -\cite{ref24}. Secondly, apart from the chaotic component inherent to the heartbeat and blood vessel motion, there exists a notable random component introduced by measurement errors in the time series under consideration.

One group of methods for distinguishing chaotic and regular series traces its origins back to the classical Rosenstein method \cite{ref25}. These methods rely on estimating the rate of divergence of initially proximate segments of the series and subsequently calculating the highest Lyapunov exponent, with positive values indicating randomness and negative values signifying regularity. Wernecke introduces a binary test designed to differentiate between strong and partially predictable chaos. Similar to the Lyapunov exponential procedure, this method depends on the evolution of initially close trajectories over time\cite{ref26}. The 0-1 test, a contemporary tool for identifying chaos in time series, comes in two forms: a regression test and a correlation test. The correlation test performs exceptionally well with noise-free time series, but when noise contaminates the data, a modified version of the original regression test minimizes sensitivity to noise \cite{ref27}. However, for real-world time series, these methods tend to exhibit a slight bias towards chaos, often producing positive values even for entirely regular series \cite{ref28},\cite{ref29}. Additionally, most methods in this category are sensitive to noise.
Another group of methods draws on the concept of ordinal structure and the representation of a time series as a point on the entropy-complexity plane \cite{ref30},\cite{ref31}. This approach discerns between complex deterministic (chaotic), simple deterministic, and stochastic time series. It rests on the observation that, unlike purely random series, deterministic series—no matter how complex and chaotic—are characterized by numerous forbidden patterns \cite{ref32},\cite{ref33}. These forbidden patterns are resilient to noise and decay at a rate contingent on the noise level, making it feasible to establish a method for distinguishing deterministic time series even in the presence of substantial noise \cite{ref4},\cite{ref34}. 

Interestingly, the absence of ordinal patterns in deterministic dynamics can indicate epileptic states \cite{ref35}. Nonetheless, real-world time series typically exhibit considerable noise levels, which curtail the applicability of this class in resolving the challenge of identifying pathological fistulas. In practical scenarios, these methods should be complemented with approaches from other classes, as demonstrated in this paper.
In the current paper, we made an attempt to address the challenge of clustering a set of time series into a predefined number of clusters, precisely two clusters. Specialists who work with hemodialysis patients have noted the need for categorizing the noises obtained from the fistula within the scope of the examined subject area. Thus, this study considers the problem of clustering time series $T_{i} = (x_{0}, x_{1}, ..., x_{N})$ into non-overlapping clusters $C_{j}: C_{j} \cap C_{k} = \emptyset, j \neq k.$ The number of clusters is not predetermined. Another task is classification, which involves assigning a given series to one of the three categories mentioned above. This procedure is based on the results of the clustering performed. 

\section{Materials and Methods}
\textbf{Methods for obtaining and describing the study sample.} During dialysis treatment, patients undergo the installation of an AVF, and the condition of the fistula is evaluated based on the continuous noise it produces. Changes in the fistula noise can indicate fistula thrombosis and the development of stenosis. The authors have developed a mobile phone application~\cite{ref36} that allows recording fistula sounds using an external microphone, along with an accumulating database of recordings~\cite{ref37}. The study involved 290 patients (131 women and 159 men) undergoing renal replacement therapy using peritoneal hemodialysis for chronic renal failure at the Dialysis Center of Surgut Regional Clinical Hospital and the Hemodialysis Department of the Samara State Medical University (Samara, Russia). The mean age of the patients was 62 (56.7 ± 5.3) years. The average duration of renal replacement therapy by renal hemodialysis was 43 (34.3 ± 8.7) months. The AVF bruit recordings were obtained from all patients for 20-30 seconds using the electronic stethoscope "Littmann Model 3200" at the auscultation point located 2 centimeters distal to the arteriovenous anastomosis. Each patient was auscultated 10 to 15 times at various times of the day before and after the hemodialysis session. In all patients, renal hemodialysis was performed three times a week through a native AVF formed in the distal third of the forearm. The research material consists of 678 records, collected from 290 patients on dialysis. The data collection and all experiments were performed in accordance with the relevant guidelines and regulations. All patients included in the study have been informed about its course and purpose, and they gave written informed consent to participate in the study. It's important to note that the information about fistula bruit for each patient was received already anonymized. The hemodialysis center doctors performed a clinical assessment of the fistulas, classifying 567 records as normal functioning, 50 as potentially problematic, and 61 as requiring imminent reconstruction. Therefore, the goal of the mathematical analysis of the spectrograms was to identify patterns that would allow for the categorization of recordings into the categories mentioned earlier.

 Each patient had from one to five records, corresponding to various states of his or her fistula, recorded at different times. We select the last segment of the audio recording of the fistula bruit (5-10 seconds) for further analysis, thereby making a time series of 1000 observations. In the course of the research, we revealed that such a length was sufficient to solve the problem in question.

\textbf{Methods for analyzing time series.} Calculation of the position on the "entropy-complexity" plane. In the work of O. A. Rosso et al. ~\cite{ref29}, a method is proposed to distinguish chaotic series from those generated by a simple deterministic system on one side, and from purely random series on the other. The approach involves computing two characteristics (entropy and complexity) for a given time series $T_{i} = \{x_{i}\}.$ The position of these characteristics on the corresponding plane relative to the upper and lower theoretical bounds determines the type of the series. The algorithm is based on constructing a probability distribution for the given series. The concept of ordinal patterns is employed to construct this distribution ~\cite{ref30}. For the constructed distribution, two characteristics are computed: entropy and complexity. 

\begin{equation}
    S[P] = -\Sigma_{i=1}^{n!} P_{i} . ln (P_{i})
\end{equation}

\begin{equation}
    H[P] = \frac{S[P]}{S_{max}}
\end{equation}

\begin{equation}
    S_{max} = ln(n!)=S[P_{e}]
\end{equation}

where $P_{e}$ is the uniform distribution $P = \{P_{i}, i = 1, 2, ..., N\}$, (Entropy achieves its maximum at this very distribution.)

We will consider the sole limiting condition for $P$, representing the state of our system, as

\begin{equation}
    \Sigma_{i=1}^{N} P_{i} = 1
\end{equation}

In the case when $S(P) = S_{min} = 0$, we can confidently predict which possible outcomes $i$ with probability $P_{i}$ will indeed occur. Thus, knowledge about the probability distribution is maximized in this case, while knowledge about the normal probability distribution $P_{e}$ is minimized:

\begin{equation}
    S_{max} = ln N = S(P_{e}),
\end{equation}

where $P_{e}$ is the uniform distribution and $N$ is the number of degrees of freedom. 
The "entropy-complexity" method is the product of the information stored in the system and its disequilibrium $H(P)$, and is computed as follows:

\begin{equation}
    C(P) = Q(P, P_{e}) . H(P),
\end{equation}

Here, $Q$ is the Jenson-Shannon divergence between the considered distribution $P$ and the uniform distribution $P_{e},$ which is calculated as follows:

\begin{equation}
    Q(P, P_{e}) = Q_{0} . (S(\frac{P+P_{e}}{2}) - S(\frac{P}{2}) - S(\frac{P_{e}}{2})),
\end{equation}

where $Q_{0}$ is the normalization constant equal to the reciprocal of the maximum possible value of $Q(P, P_{e}),$ computed by the formula:

\begin{equation}
    Q_{0} = -2\{(\frac{N+1}{N}) ln(N+1) - 2 ln (2N) + lnN\}^{-1}
\end{equation}

It should be noted that the statistical complexity defined above is the product of two normalized entropies (Shannon entropy and Jenson-Shannon divergence). However, it is a non-trivial function of entropy since it depends on two different probability distributions, namely, the one corresponding to the system state $P$ and the uniform distribution $P_{e}.$ In the case of the analyzed series in this work, the complexity is always positive, $C > 0.$ It is worth noting that the value of complexity varies depending on the nature of the observations and the size of the series used. Depending on the size of the series, each observation introduces a new set of available states with their corresponding probability distribution. The procedure described above maps each series to a point on the "entropy-complexity" plane, and the position of that point relative to the upper and lower theoretical bounds indicates the type of the observed series ~\cite{ref29}.

\textbf{The Wishart clustering algorithm.} To solve the given problem, it is necessary to perform clustering on the specified dataset. The primary requirement for selecting a clustering algorithm is the absence of prior information about the number of clusters ~\cite{ref38}. Thus, in this study, the clustering method proposed by Wishart ~\cite{ref39, ref40}, modified by Lapko and Chentsov ~\cite{ref41} for cluster vectors, was used. Some difficulties associated with applying the algorithm to time series are discussed in ~\cite{ref42}. 

\textbf{Construction of the "objects-concepts" graph and its analysis.} Within the framework of formal concept analysis ~\cite{ref18}, an alternative approach to distinguishing regular series from chaotic ones was proposed. It involves constructing an "objects-concepts" graph ~\cite{ref19}~\cite{ref20}. Such graphs comprise vertices of two types: vertices corresponding to the set of the objects, and vertices containing subsets of attributes that characterise these objects. In particular, community detection algorithms on these graphs make it possible to obtain not mere clusters, but clusters and sets of attributes that determine these clusters.
Series with pronounced chaos yield graphs with an essentially more complex structure than those of graphs corresponding to regular time series. To evaluate the degree of structural complexity, we employed the following statistics: the average number of elements per cluster; and the share of non-cluster noise. To construct the clusters, we will use the asynchronous label propagation algorithm described earlier ~\cite{ref21}. Figure 1b shows the result of applying this algorithm to the graph in Figure 1a.

\begin{figure}
\centering
\includegraphics[width=1\columnwidth]{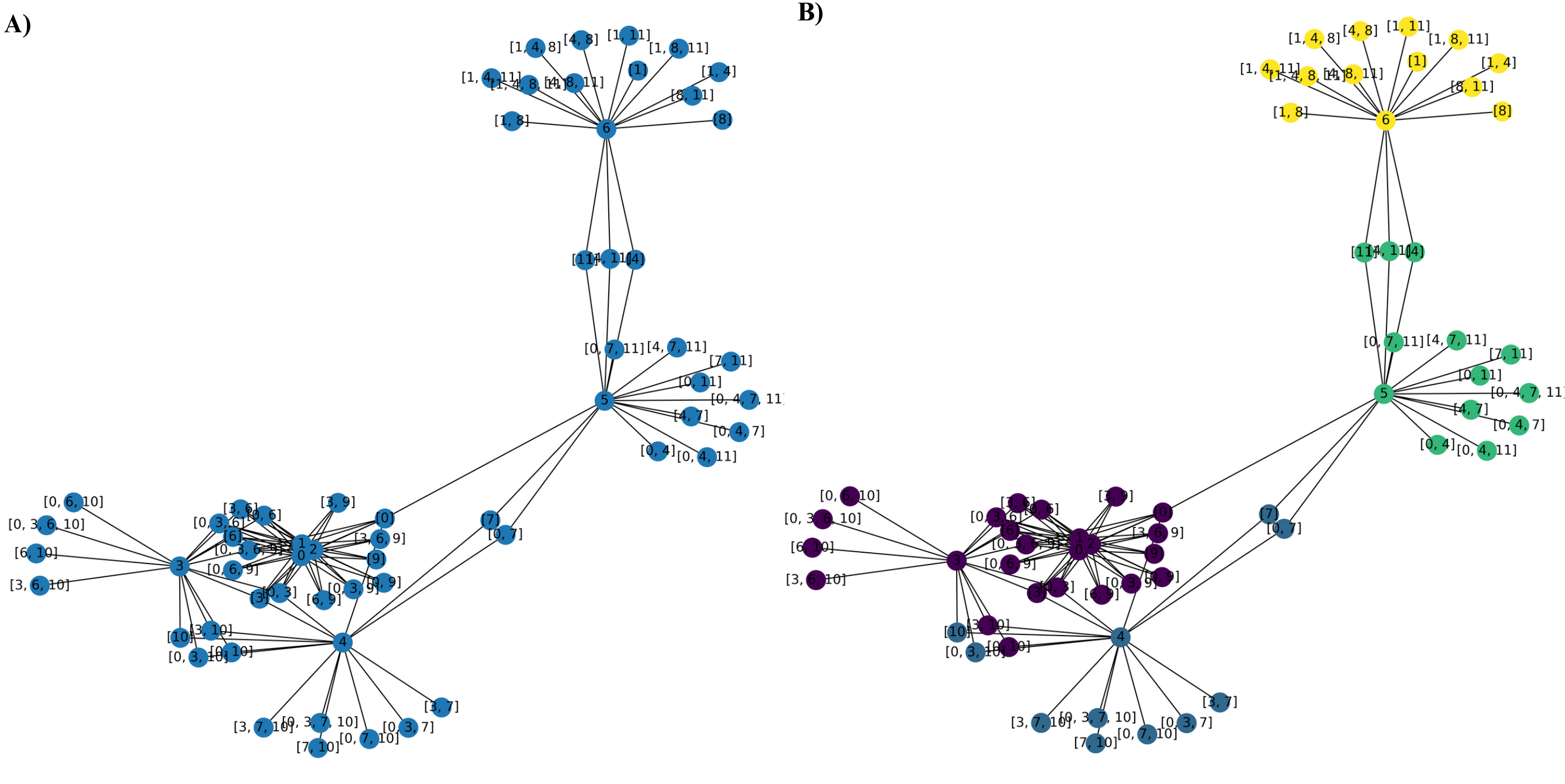}
\caption{Communities detected by the asynchronous label propagation (B) in the concepts-objects graph (A).}
\end{figure}

To assess the level of structural complexity of the graph, the following statistics were formulated:

\begin{itemize}
    \item The number of clusters relative to the total number of elements.
    \item The proportion of non-cluster noise.
    \item In the course of a large-scale simulation, we found that the effective boundaries between regular and chaotic series are $10\%$ for both the first statistics and the second.
statistics.
\end{itemize}

\section{Results}
Figure 2a shows a typical 15-second sequence for a patient with a normally functioning fistula, while Figure 2b shows a sequence for a patient with fistula dysfunction. For comparison, Figure 2c presents a segment of the Lorenz series, which is a typical chaotic series.

\begin{figure}
\centering
\includegraphics[width=0.68\columnwidth]{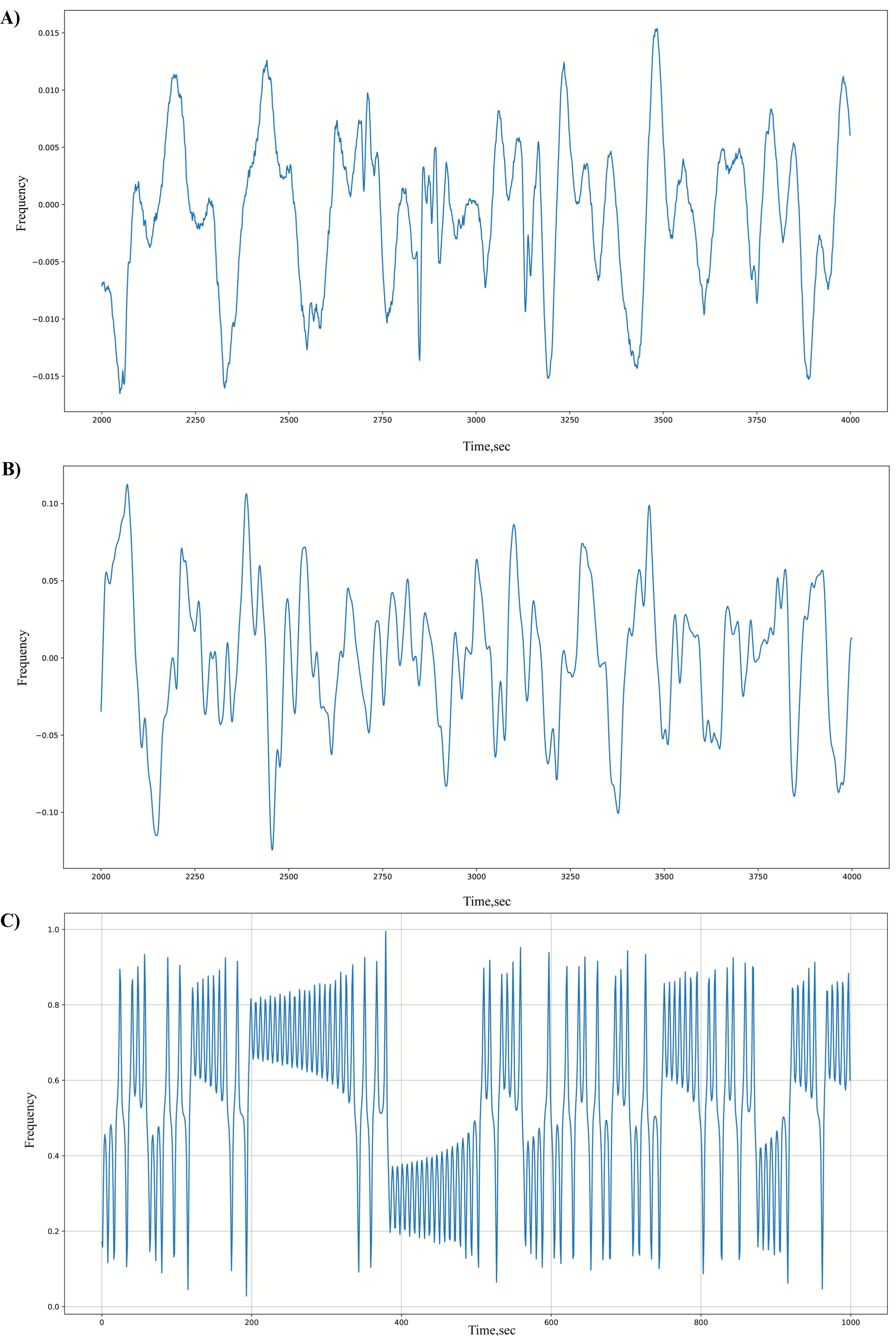}
\caption{15-second recordings for a patient with a normally working fistula (A); with fistula dysfunction (B); sequence corresponding to the Lorenz series (C).}
\end{figure}

For each time series element in the examined dataset, the algorithm for constructing a point on the "entropy-complexity" plane was applied. Subsequently, the obtained dataset was subjected to the Wishart clustering algorithm, which does not require prior knowledge of the number of clusters and determines it autonomously during the clustering process (Figure 3).

\begin{figure}
\centering
\includegraphics[width=0.8\columnwidth]{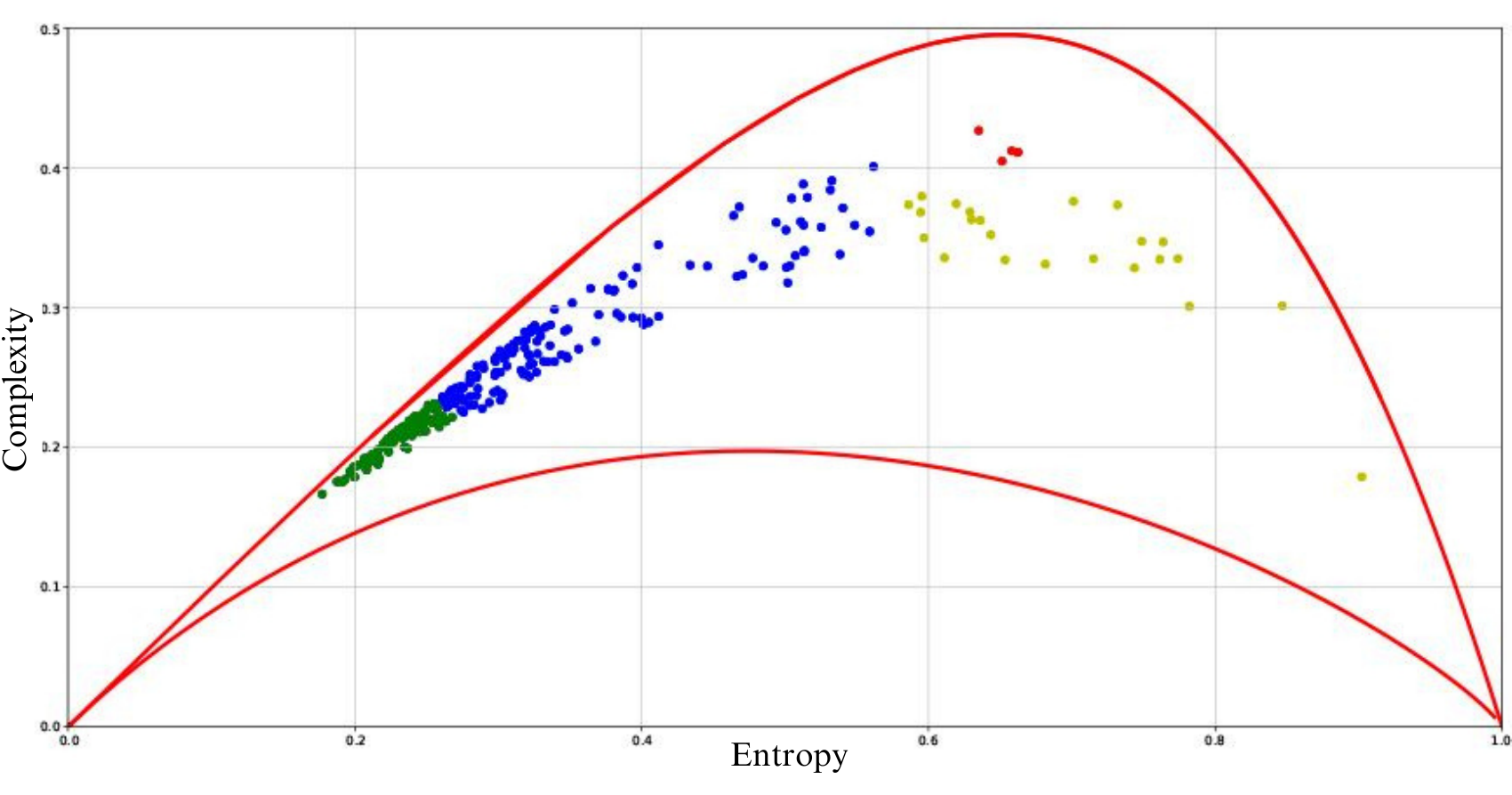}
\caption{Wishart algorithm clusterisation results.}
\end{figure}

The clustering algorithm identified three clusters:
\begin{itemize}
    \item The first one (green color in the figure) lies in the region of regular movements and, according to our hypothesis, corresponds to normally functioning fistulas.
    \item The second cluster (yellow color) is in the region of chaotic movements and movements close to chaos, corresponding to fistulas with dysfunction. It contains a small red subcluster, which represents samples with abnormal external noises.
    \item the third cluster (blue color) occupies an intermediate position and corresponds to uncertain patients. An alternative approach in analyzing the considered time series is the construction of an "objects-concepts" graph with the identification of clusters based on communities within the graph.
\end{itemize}

Figure 4a shows characteristic graphs corresponding to the case of a normally functioning fistula, while Figure 4b shows the case of fistula dysfunction obtained with the FCA method. To obtain integrated results using both the first and second approaches, a linear regression method was employed. Even without resorting to formal methods, we may observe that the graphs are distinctly different. For the first graph, the average number of elements per cluster amount is 0.091, the percentage of extra-cluster noise is $2.4\%$, maximal cluster size is 0.149. For the second graph, the average number of elements per cluster amount is 0.329, the extra-cluster noise is  $28.7\%$ and the maximal cluster size is 0.283 respectively.

The Wishart clustering algorithm was applied to the values of the metrics thus obtained. The objects-concepts algorithm distinguishes only two classes, corresponding to the states of - ‘fistula works well’, and ‘there’s a slight malfunction in the fistula’. In order to integrate results of the first and second approaches, we use linear regression. The integrated results appear to be more accurate in identifying fistula dysfunction than the methods alone. The integrated results were more accurate in determining fistula dysfunction compared to applying each method individually.

\begin{figure}
\centering
\includegraphics[width=1\columnwidth]{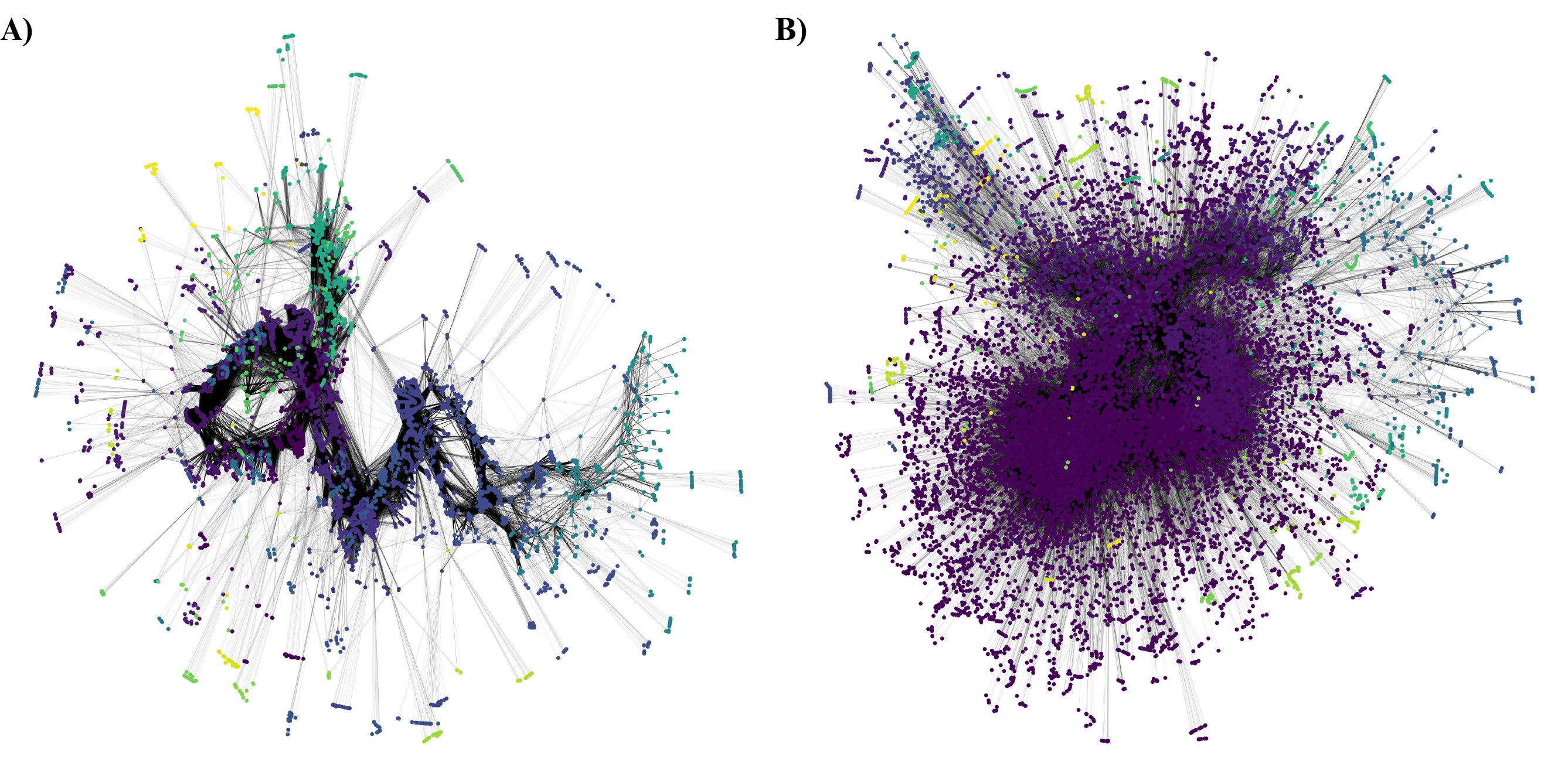}
\caption{(A)-Characteristic graph corresponding to a case of a normally functioning fistula (number of clusters - 45, proportion of noise - 0.024). (B)-Characteristic graph corresponding to a case of fistula dysfunction (number of clusters - 163, proportion of noise - 0.287).}
\end{figure}

To verify the obtained results, a comparison was made with the solutions adopted by the attending physicians regarding the patients' data. The table presents the mentioned comparisons for the analysis conducted within the framework of the first and second approaches, as well as the integrated results.

\begin{table}[]
\centering
\caption{\label{demo-table} Relative efficiency of the approaches on the sample used (n = 290).}
\begin{tabular}{lllll}
\hline
Method                                                                                                                                                      & \begin{tabular}[c]{@{}l@{}}True \\ Positive\end{tabular} & \begin{tabular}[c]{@{}l@{}}True \\ Negative\end{tabular} & \begin{tabular}[c]{@{}l@{}}False \\ Positive\end{tabular} & \begin{tabular}[c]{@{}l@{}}False \\ Negative\end{tabular} \\ \hline
\begin{tabular}[c]{@{}l@{}}Entropy-Complexity (Wishart):\\ Precision = 0.76\\ Recall = 0.81\\ F-measure = 0.878\\ Accuracy = 72.3\end{tabular}              & 130                                                      & 90                                                       & 40                                                        & 30                                                        \\ \hline
\begin{tabular}[c]{@{}l@{}}objects-concepts graphs:\\ Precision = 0.76\\ Recall = 0.78\\ F-measure = 0.77\\ Accuracy = 72.3\end{tabular}                    & 140                                                      & 100                                                      & 30                                                        & 20                                                        \\ \hline
\begin{tabular}[c]{@{}l@{}}Linear regression based on both algorithm:\\ Precision = 0.90\\ Recall = 0.77\\ F-measure = 0.83\\ Accuracy = 84.50\end{tabular} & 170                                                      & 70                                                       & 20                                                        & 30                                                        \\ \hline
\end{tabular}
\end{table}

We should stress that the diagnostic decisions of doctors (with which we compare our results) should not be considered as the ultimate truth: the overwhelming majority of cases of false positive and false negative diagnoses were related to situations when the doctors are not sure of their decision. They found the cases ambiguous and requiring additional analysis.

\section{Conclusions and future directions}
The study employed a methodology to automatically distinguish between series representing normally and pathologically functioning arteriovenous fistulas (AVFs). This method relies on computing a pair of characteristics known as "entropy-complexity." Additionally, a novel approach was introduced, involving the creation of an objects-concepts graph for series classification and an analysis of its properties. Notably, these properties exhibited qualitative differences between regular and chaotic series. Both of these approaches, when used individually or in combination, were applied to assess the condition of AVFs and demonstrated high diagnostic effectiveness.

A promising avenue for future research in this field, particularly relevant to healthcare practitioners, could involve a thorough exploration of the structure of the objects-concepts graph to identify characteristic substructures associated with various types of fistula dysfunction.

In previous research, various classification methods were employed to categorize arteriovenous fistulas (AVFs) into distinct subclasses. Some authors utilized acoustic frequency domain analysis, employing the Fast Fourier Transform (FFT) and subsequent clustering \cite{ref15}, \cite{ref43}, \cite{ref15}, resulting in the introduction of six subclasses. Simultaneously, prior studies introduced a deep learning method that defined five AVF subclasses ~\cite{ref16}, ~\cite{ref44}, ~\cite{ref45}. However, these methods were found to be sensitive to the quality of the recorded acoustic data.

In contrast, our proposed method exhibits reduced dependence on noise extraction from the records. The first algorithm reveals three clusters, corresponding to healthy, malfunctioning, and intermediate fistulas. The second algorithm distinguishes four classes, including bad and extremely bad among malfunctioning fistulas. The Wishart clusterization method, which we employ, does not impose constraints on the number of clusters, allowing us to explore beyond the originally defined three states of the fistula. Moreover, we have observed a significant correlation between the classification results obtained using our methods and the outcomes of traditional diagnostic tests commonly employed to evaluate AVF function, such as Doppler ultrasound.

It is crucial to acknowledge the limitations of our study, including the relatively small sample size of patients studied and the subjective classification of AVFs into functional categories, based on the expertise of clinicians in the dialysis unit.

As a possible direction for future research, we may indicate that the structure of the concepts-objects graph allows identifying typical substructures corresponding to various variants of fistula dysfunction, which is important from the point of view of practitioners. We also plan to design algorithms for the early diagnosis of fistula health, which will assess the chances of a healthy functioning fistula within the first or second day after the operation.

\subsubsection{Acknowledgements} 
The authors would like to acknowledge that this paper is the result of a research project conducted under the auspices of the Basic Research Program at the National Research University Higher School of Economics (HSE University).

\end{document}